\documentclass[12pt]{article}
\usepackage[english]{babel}
\usepackage{amssymb,slashed,latexsym,amsmath,multirow,color}
\usepackage{slashed}
\pdfoutput=1
\usepackage[font=footnotesize,labelsep=newline,labelfont=sc]{caption}

\numberwithin{equation}{section}

\begin{document}
\begin{titlepage}
\begin{flushright}

\end{flushright}
\vspace{.5cm}
\begin{center}
\baselineskip=16pt {\LARGE    Symplectic gaugings and the field-antifield formalism
}\\
\vfill
{\large Frederik Coomans$^{\dag}$, Jan De Rydt$^{\dag}$, Antoine Van Proeyen$^{\dag}$
  } \\
\vfill
{\small Instituut voor Theoretische Fysica, Katholieke Universiteit Leuven,\\
       Celestijnenlaan 200D B-3001 Leuven, Belgium.
\\ \vspace{6pt}
 }
\end{center}
\vfill
\begin{center}
{\bf Abstract}
\end{center}
{\small 
We give an example of how conventional gauging methods obstruct a systematic analysis of gauged supergravities. We discuss how the embedding tensor formalism deals with these problems and argue that the gauge algebra related to the embedding tensor formalism is soft, open and reducible. We connect the embedding tensor formalism to the field-antifield (or Batalin-Vilkovisky) formalism, which is the most general formulation known for gauge theories. 
}
 \vfill

\hrule width 3.cm \vspace{2mm}{\footnotesize \noindent $^{\dag}$e-mails:
\{Frederik.Coomans, Jan.DeRydt, Antoine.VanProeyen\}@fys.kuleuven.be}
\end{titlepage}
\addtocounter{page}{1}
 
\section{Introduction}

This text is based on a talk given by F.C. at the XVIth European Workshop on String Theory in Madrid and on \cite{Coomans:2010xd}. 

Properties of supergravity theories depend primarily on the choice of spacetime dimension $D$ and the number of gravitini ${\cal N}$. There are basic supergravities and deformed supergravities. Basic theories have no potential or mass terms in their Lagrangian. They are determined by supersymmetry and the kinetic terms. Basic theories with more than 16 supercharges have unique kinetic terms for every dimension $D$. Also the theories with 16 supercharges are completely fixed once $D$ and the number of vector (or tensor) multiplets coupled to supergravity are given. Basic theories with less than 16 supercharges have more freedom. Here one still has the choice of adding different matter multiplets. In these cases the model also depends on some functions that can vary by infinitesimal variations (e.g. a K\"ahler potential or a prepotential). All in all the basic supergravities are well understood and classified.

After specifying $D$, ${\cal N}$ and (eventually) the matter multiplets there are still different theories possible. These are deformations of the basic theories described above. The simplest example of a deformation is the choice of a superpotential in $D=4$, ${\cal N}=1$ supergravity. Other examples consist in the coupling to Yang-Mills type gauge groups (see for example \cite{de_Wit:1982ig}) or the introduction of mass-parameters \cite{Romans:1985tz}. In this text we are concerned about the coupling to Yang-Mills type gauge groups, the so-called gauged supergravities. In the past ten years a powerful technique has been developed to systematically study and classify these theories. This technique is the embedding tensor formalism \cite{Cordaro:1998tx,Nicolai:2000sc,Nicolai:2001sv,deWit:2002vt,deWit:2004nw,deWit:2005hv,deWit:2005ub}. If we want to obtain a full classification of gauged supergravities it is necessary to thoroughly study the structure of this embedding tensor formalism. 

In this text we will first give a short introduction to this formalism by applying it to a generic $D=4$, ${\cal N}=1$ supergravity theory. Then we will study the structure of the resulting gauge algebra and conclude that it is soft, open and reducible. Since these are precisely the features of the field-antifield formalism of Batalin and Vilkovisky\footnote{For an introduction consult \cite{Gomis:1995he,Henneaux:1990jq}.} \cite{Batalin:1984jr}, it is natural to use it to reformulate the embedding tensor formalism. This will be discussed in the last part of the text.   

\section{$D=4$, ${\cal N}=1$ supergravity and conventional gauging}

A generic ungauged $D=4$, ${\cal N}=1$ supergravity has the following bosonic Lagrangian for its gauge and scalar fields
\begin{eqnarray}
e^{-1}\mathcal{L}_{\rm{s}}+e^{-1}\mathcal{L}_{\rm{g.k.}}&=&-g_{ \alpha \bar{\beta}}\partial_{\mu}z^{\alpha}\partial^{\mu}\bar{z}^{\bar{\beta}}-\frac{1}{4} \text{Re} f_{\Sigma\Lambda}(z) F_{\mu\nu}{}^{\Sigma}F^{\mu\nu}{}^{\Lambda} \nonumber \\
&&+\frac{1}{8}e^{-1}\text{Im} f_{\Sigma\Lambda}(z)\varepsilon^{\mu\nu\rho\sigma}F_{\mu\nu}{}^{\Sigma}F_{\rho\sigma}{}^{\Lambda}. \label{bosoniclagrangian}
\end{eqnarray}
The $f_{\Sigma\Lambda}(z)$ are scalar dependent positive definite functions and the $F_{\mu\nu}{}^{\Lambda}$ (with $\Lambda=1, \ldots, n$) are the (up to now) abelian field strengths. $n$ is the number of vector fields. The scalar fields $z^{\alpha}$ parametrize a K\"ahler manifold with $g_{\alpha \bar{\beta}}=\partial_{\alpha}\partial_{\bar{\beta}}K(z,\bar{z})$ the K\"ahler metric and $K(z,\bar{z})$ the K\"ahler potential.
There are no gauge couplings and no scalar potential. This Lagrangian has only global symmetries and abelian gauge groups. The theory is determined completely by the gauge kinetic function $f_{\Sigma\Lambda}(z)$ and the K\"ahler potential $K(z, \bar{z})$.

Before we can switch on any gauge couplings, we need to determine the rigid (global) symmetry group of the theory $G_{\rm{rigid}}$. This is the group of global transformations that leave the total set of Bianchi identities and equations of motion of the theory unchanged but, in general, do not leave the Lagrangian invariant\footnote{$G_{\rm{rigid}}$ is also denoted as the group of electric/magnetic duality transformations.}. We know that the rigid symmetry group of the vector sector is $\rm{Sp}(2n,\mathbb{R})$. On the other hand, the rigid symmetry group of the scalar sector is given by the isometry group of the scalar manifold $\rm{Iso}(\mathcal{M}_{\rm{scalar}})$. Hence we conclude that, since the scalar and vector sectors are coupled via the gauge kinetic functions $f_{\Sigma\Lambda}(z)$, the rigid symmetry group of the full theory (scalar and vector sector)
\begin{equation}
G_{\rm{rigid}} \subseteq \rm{Sp}(2n,\mathbb{R}) \times \rm{Iso}(\mathcal{M}_{\rm{scalar}}).
\end{equation}
The rigid symmetry group is determined by the gauge kinetic function $f_{\Sigma\Lambda}(z)$ and the K\"ahler potential $K(z,\bar{z})$. 

As already mentioned above, $G_{\rm{rigid}}$ does not necessarily leave the Lagrangian invariant. Only a subgroup of $G_{\rm{rigid}}$ leaves the Lagrangian unchanged and only this subgroup (or a subgroup thereof) can, according to the method of conventional gauging, be promoted to a gauge group. The method of conventional gauging is the following. Let us denote the generators of $G_{\rm{rigid}}$ as $\delta_a$, where $[\delta_a,\delta_b]=f_{ab}{}^c\delta_c$ with $f_{ab}{}^c$ the structure constants of $G_{\rm{rigid}}$. We select a gauge group $G_{\rm{gauge}} \subseteq G_{\rm{rigid}}$ by choosing linear combinations of the $\delta_a$
\begin{equation}
\delta_{\Lambda}=\Theta_{\Lambda}{}^a\delta_a
\end{equation}
that leave the Lagrangian invariant. $\Theta_{\Lambda}{}^a$ is called the embedding tensor that determines the embedding of $G_{\rm{gauge}}$ into $G_{\rm{rigid}}$ and that satisfies a constraint ensuring closure of the gauge generators. We then promote the symmetries $\delta_{\Lambda}$ to local ones by switching on gauge couplings in covariant derivatives and non-abelian field strengths
\begin{equation}
D_{\mu}=\partial_{\mu}-gA_{\mu}{}^{\Lambda}\delta_{\Lambda}, \qquad \qquad F_{\mu\nu}{}^{\Lambda}=2\partial_{[\mu}A_{\nu]}{}^{\Lambda}+gf_{\Sigma\Omega}{}^{\Lambda}A_{\mu}{}^{\Sigma}A_{\nu}{}^{\Omega}, \label{covder}
\end{equation}
where $f_{\Sigma\Omega}{}^{\Lambda}$ are the structure constants of $G_{\rm{gauge}}$. 

This method of conventional gauging brings along several problems obstructing our aim to systematically study all possible gaugings of the theory. First of all only electric couplings are possible (because only electric gauge fields $A_{\mu}{}^{\Lambda}$ appear in the covariant derivatives (\ref{covder})). Hence, this conventional gauging method already breaks the duality covariance of the set of field equations and Bianchi identities. The second problem is that the possible gauge groups depend on the symplectic frame since only subgroups of $G_{\rm{rigid}}$ that leave the Lagrangian invariant can be promoted to a gauge group. All this asks for a more general gauging mechanism. This will be provided by the embedding tensor formalism.

\section{The embedding tensor formalism}
To solve the problems mentioned at the end of the previous section and hence provide a duality covariant method of gauging the 4 dimensional Lagrangian (\ref{bosoniclagrangian}), one introduces magnetic vector fields $A_{\mu}{}_{\Lambda}$ in addition to the usual electric ones $A_{\mu}{}^{\Lambda}$. The $A_{\mu}{}_{\Lambda}$ are defined as follows
\begin{equation}
G_{\mu\nu}{}_{\Lambda}=2\partial_{[\mu}A_{\nu]}{}_{\Lambda} \qquad \text{and} \qquad G_{\mu\nu}{}_{\Lambda} \equiv \varepsilon_{\mu\nu\rho\sigma}\frac{\delta \mathcal{L}}{\delta F_{\rho\sigma}{}^{\Lambda}}.
\end{equation}
These magnetic vectors are combined with the electric vectors in a symplectic vector 
\begin{equation}
A_{\mu}{}^M=\Bigl(A_{\mu}{}^{\Lambda},A_{\mu}{}_{\Lambda}\Bigr).
\end{equation}
Under the rigid symmetry group, the vectors transform as follows
\begin{equation}
\delta(\lambda)A_{\mu}{}^M=\lambda^a\delta_aA_{\mu}{}^M=-\lambda^a(t_a)_N{}^MA_{\mu}{}^N,
\end{equation}
where the $(t_a)_N{}^M$ are the rigid symmetry generators in the vector representation\footnote{The matrices $(t_a)_N{}^M$ satisfy the symplectic condition $(t_a)_{(N}{}^M\Omega_{P)M}=0$, where $\Omega_{PM}$ is the symplectic metric.}. Also the embedding tensor is generalized to
\begin{equation}
\Theta_M{}^a=\Bigl(\Theta_{\Lambda}{}^a,\Theta^{\Lambda}{}^a\Bigr).
\end{equation}
Once the gauge generators $\delta_M$ have been selected through the choice of an embedding tensor $\delta_M=\Theta_M{}^a\delta_a$, one proceeds as usual by introducing a local parameter, $\Lambda^M(x)$, for every generator and by constructing covariant derivatives. The gauge transformations of the vectors are
\begin{equation}
\delta(\Lambda)A_{\mu}{}^M=\partial_{\mu}\Lambda^M+gA_{\mu}{}^NX_{NP}{}^M\Lambda^P 
\end{equation}
with $X_{NP}{}^M \equiv \Theta_N{}^a(t_a)_P{}^M$. Covariant derivatives are defined as
\begin{equation}
D_{\mu}=\partial_{\mu}-gA_{\mu}{}^{M}\delta_{M}=\partial_{\mu}-gA_{\mu}{}^{\Lambda}\Theta_{\Lambda}{}^a\delta_a-gA_{\mu}{}_{\Lambda}\Theta^{\Lambda}{}^a\delta_a. \label{emcovder}
\end{equation}
The covariant derivatives can now contain electric \textit{and} magnetic gauge fields, hence keeping the duality covariance of the ungauged theory.

Of course this duality covariance does not come for free. A consistent gauging requires the imposition of a closure constraint on the embedding tensor. It has the following form
\begin{equation}
f_{bc}{}^a\Theta_M{}^b\Theta_{N}{}^c+(t_b)_N{}^Q\Theta_M{}^b\Theta_Q{}^a=0.
\end{equation}
The definition of $X_{NM}{}^P$ together with the closure constraint implies 
\begin{equation}
[X_M,X_N]=-X_{MN}{}^PX_P, \label{Xcommrel}
\end{equation} 
which can be interpreted as closure of the gauge algebra with $X_{MN}{}^P$ as its generalized structure constants. Equation (\ref{Xcommrel}) implies that $X_{(NM)}{}^P$ only vanishes upon contraction with the embedding tensor, but, in general, is not zero in itself. This signals a difference with ordinary gauge groups, where the structure constants are antisymmetric and satisfy a Jacobi identity. In our case the Jacobi identity is violated\footnote{Instead a generalized Jacobi identity is valid: $$ X_{QR}{}^MX_{(NP)}{}^R-X_{(RP)}{}^MX_{QN}{}^R-X_{(NR)}{}^MX_{QP}{}^R=0.$$ \label{footnoteJacobi}} by terms that are proportional to $X_{(NM)}{}^P$. 

In order to maintain gauge covariance of the theory it is required to introduce several new ingredients (which will seriously complicate the gauge structure of the theory). We will now, very schematically, introduce these necessary ingredients. For a more thorough discussion we refer to \cite{deWit:2005ub, deWit:2008ta, Coomans:2010xd}. 
\begin{itemize}
\item The first step towards gauge covariance is the introduction of extra gauge transformations for the vector fields,
accompanied by new local parameters $\Xi_\mu{}^{NP}(x)$:
\begin{equation}\label{deltaA}
 \delta(\Lambda) A_\mu{}^M\;\rightarrow\;\delta(\Lambda,\Xi) A_\mu{}^M = \delta(\Lambda) A_\mu{}^M + \delta(\Xi) A_\mu{}^M, 
\end{equation}
where
\begin{equation}
\delta(\Xi) A_\mu{}^M = - X_{(NP)}{}^M \Xi_\mu{}^{NP}\,.
\end{equation}
These new transformations, proportional to $X_{(MN)}{}^P$, are introduced to gauge away the directions in the algebra that violate the Jacobi identity.
\item The next step in the construction of the theory is the introduction of covariant field strengths for the vectors. The usual expression, ${\cal F}_{\mu\nu}{}^M=2\partial_{[\mu}A_{\nu]}{}^M+X_{[NP]}{}^MA_{\mu}{}^NA_{\nu}{}^P$, does not transform covariantly but picks up terms that are proportional to $X_{(NP)}{}^M$. Therefore, we will introduce new field strengths,
\begin{equation}\label{covH}
 {\cal H}_{\mu\nu}{}^M\equiv {\cal F}_{\mu\nu}{}^M+X_{(NP)}{}^MB_{\mu\nu}{}^{NP}\,,
\end{equation}
and new $2$-forms $B_{\mu\nu}{}^{NP}$. The gauge transformations of these 2-forms are fixed by demanding the covariant transformation of the new field strengths ${\cal H}_{\mu\nu}{}^M$
\begin{equation}\label{deltaB}
 \delta(\Lambda,\Xi) B_{\mu\nu}{}^{NP} = 2 {D}_{[\mu}\Xi_{\nu]}{}^{NP} + 2 A_{[\mu}{}^{(N}\delta A_{\nu]}{}^{P)} - 2 \Lambda^{(N}{\cal H}_{\mu\nu}{}^{P)}.
\end{equation}
 \item Realising this duality covariant gauging method at the level of the action (\ref{bosoniclagrangian}) requires several modifications. First of all the field strengths $F_{\mu\nu}{}^{\Lambda}$ in $\mathcal{L}_{\rm{g.k.}}$ need to be replaced by the generalized ones ${\cal H}_{\mu\nu}{}^{\Lambda}$ and the derivatives in $\mathcal{L}_{\rm{s}}$ need to be replaced by the covariant ones (\ref{emcovder}). Gauge invariance also requires the addition of Chern-Simons terms for the vectors, denoted by $\mathcal{L}_{\rm{GCS}}$, and topological terms for the 2-forms, denoted by $\mathcal{L}_{\rm{top}, B}$. We will denote the full Lagrangian as
\begin{equation}
\mathcal{L}_0=\mathcal{L}_{\rm{g.k.}}+\mathcal{L}_{\rm{GCS}}+\mathcal{L}_{\rm{top}, B}+\mathcal{L}_{\rm{s}}. \label{action}
\end{equation}
The full expression for this modified Lagrangian can be found in \cite{deWit:2005ub}. In \cite{deWit:2005ub, DeRydt:2008hw} it was pointed out that, in the presence of an action, (\ref{deltaB}) needs to be modified to
\begin{equation}
 \delta(\Lambda,\Xi) B_{\mu\nu}{}^{NP} = 2 {D}_{[\mu}\Xi_{\nu]}{}^{NP} + 2 A_{[\mu}{}^{(N}\delta A_{\nu]}{}^{P)} - 2 \Lambda^{(N}{\cal G}_{\mu\nu}{}^{P)}, \label{Btransform}
\end{equation}
where ${\cal G}_{\mu\nu}{}^M =({\cal H}_{\mu\nu}{}^\Lambda,{\cal G}_{\mu\nu\Lambda})$, with ${\mathcal G}_{\mu\nu\,\Lambda} \equiv \varepsilon_{\mu\nu\rho\sigma}\frac{\partial {\cal L}}{\partial {\cal H}_{\rho\sigma}{}^\Lambda}$.
\end{itemize}
To summarize, we have an action (\ref{action}) that is invariant\footnote{Invariance requires also a second constraint to be imposed on the embedding tensor. This is the so-called linear constraint: $X_{(MN}{}^P\Omega_{Q)P}=0$.} under the gauge transformations (\ref{deltaA}) and (\ref{Btransform}) and some nonlinear transformations of the scalars which are not important in the rest of our discussion.

\section{Structure of the gauge algebra}
From the previous section it is clear that there is a price we pay for the duality covariance of the gauged theory: the gauge structure becomes increasingly complicated. We had to introduce new types of gauge transformations and new 2-form gauge fields. Had we chosen a theory with $D > 4$ to illustrate the embedding tensor formalism, then the gauge structure would become even more complicated. More higher-form gauge fields and hence more types of gauge transformations would have to be introduced to ensure gauge covariance. An explicit discussion about this \textit{hierarchy} of form-fields and gauge transformations can be found in \cite{deWit:2008ta, deWit:2008gc, Bergshoeff:2009ph}.

In this section we will discuss the structure of the gauge algebra of the ${\cal N}=1$, $D=4$ theory. We expect that this structure can be extended to theories with a more complex tensor hierarchy. For an explicit discussion see \cite{Coomans:2010xd}.
\begin{itemize}
 \item The gauge algebra is \textbf{open}: the commutator of two gauge transformations\footnote{Closure of the gauge algebra also requires a third type of gauge transformations ($\Phi$-type), only working on the two-forms.} ($\Lambda$-, $\Xi$- and/or $\Phi$- type) contains, in addition to a linear combination of transformations, terms that are proportional to the field equations (from the Lagrangian (\ref{action})). 
 \item The gauge algebra is \textbf{soft}: the structure `constants' are field-dependent. In our case, the non-zero commutators are $f_{\Lambda\Lambda}{}^{\Lambda}$, $f_{\Xi\Xi}{}^{\Phi}$, $f_{\Lambda\Lambda}{}^{\Xi}$, $f_{\Lambda\Lambda}{}^{\Phi}$, $f_{\Lambda\Xi}{}^{\Phi}$, from which the latter three are all field dependent.
 \item The gauge algebra is \textbf{reducible}: the different gauge transformations are not all independent, there are zeromodes. Let us give an example here. The gauge transformation of the vector $A_{\mu}{}^M$ is
\begin{equation}
\delta A_{\mu}{}^M=\partial_{\mu}\Lambda^M+A_{\mu}{}^NX_{NP}{}^M\Lambda^P-X_{(NP)}{}^M\Xi_{\mu}^{NP}. \label{fullAtransform}
\end{equation}
This transformation vanishes for an appropriate choice of the parameters $\Lambda$ and $\Xi$ :
\begin{equation}
\Lambda^M=X_{(RS)}{}^M\Lambda^{RS} \qquad \text{and} \qquad \Xi_{\mu}{}^{NP}=\partial_{\mu}\Lambda^{NP}+2A_{\mu}{}^QX_{QR}{}^{(N}\Lambda^{P)R},
\end{equation} 
with\footnote{To prove the vanishing of the transformation, the generalized Jacobi identity, mentioned in footnote \ref{footnoteJacobi}, needs to be used.} $\Lambda^{RS}=\Lambda^{SR}$ an arbitrary local parameter. This specific choice of the parameters $\Lambda^M$ and $\Xi_{\mu}{}^{NP}$ corresponds to a zero mode of the theory which is parametrized by $\Lambda^{RS}$. 
  \item The gauge algebra is also \textbf{higher stage reducible}: the different zeromodes are not all independent. There are zeromodes for the zeromodes. These second stage zeromodes also turn out not to be independent: there are again zeromodes for these second stage zeromodes and so on.
\end{itemize}
Hence, the gauge algebra related to the embedding tensor formalism is soft, open and reducible. These are precisely the features for which the field-antifield formalism \cite{Batalin:1984jr} was designed. The original aim of this formalism was to provide a method to quantize general gauge theories but - and this is the main reason why it is used in \cite{Coomans:2010xd} - it also provides a very concise way to write down the gauge structure of a classical theory into one \textit{master equation}. 

\section{The field-antifield formalism}
The main ingredients of the field-antifield formalism are the following
\begin{itemize}
 \item The `classical' fields $\phi^i \equiv \{A_{\mu}{}^M, B_{\mu\nu}{}^{MN}\}$ are completed with ghosts $c_{(0)}^{a_0}$ (corresponding to the gauge transformations), ghost for ghosts $c_{(1)}{}^{a_1}$ (corresponding to the first stage zeromodes) and so on (see table \ref{hierarchy}). Let us call all these fields $\chi^n \equiv \{\phi^i,c_{(0)}{}^{a_0},c_{(1)}{}^{a_1},\ldots\}$. For all the fields $\chi^n$ also an additional antifield $\chi^*_n$ is introduced.
 \begin{table}[ht]
\centering
\begin{tabular}{|l|lll|}
    \hline fields $\phi^i$&&$A_\mu{}^M$&$B_{\mu\nu}{}^{ MN }$\\
ghosts $c_{(0)}{}^{a_0}$ & $c_{(0)}{}^{M}$&$c_{(0)}{}_\mu{}^{MN}$&$c_{(0)}{}_{\mu\nu}{}^{MNP}$\\
$c_{(1)}{}^{a_1}$&$c_{(1)}{}^{MN}$&$c_{(1)}{}_{\mu}{}^{MNP}$&$c_{(1)}{}_{\mu\nu}{}^{MNPQ}$\\
\multicolumn{1}{|c|}{$\vdots$}&\multicolumn{1}{|c}{$\vdots$}&\multicolumn{1}{c}{$\vdots$}&\multicolumn{1}{c|}{$\vdots$}\\ \hline
  \end{tabular}
\caption{Hierarchies at the level of classical fields, ghosts and ghost for ghosts. \label{hierarchy}}
\end{table}
\item The action $S_0=S_0[\phi^i]$ (from (\ref{action})) is extended to a generalized action $S[\chi^n,\chi^*_n]$ also containing terms proportional to ghosts and antifields and subject to some boundary conditions\footnote{These are: the classical limit $S[\chi^n,\chi^*_n=0]=S_0[\phi^i]$ and a properness condition (see \cite{Batalin:1984jr,Gomis:1995he}).}.
  \item The classical master equation is imposed 
\begin{equation}
(S,S) \equiv 2\frac{\delta_r S}{\delta \chi^n}\frac{\delta_l S}{\delta \chi^*_n}=0, \label{masterequation}
\end{equation}
where $\delta_r$ and $\delta_l$ mean right and left derivative respectively. Solutions of this master equation (subject to the boundary conditions mentioned above) are an expansion in the antifields with the gauge structure tensors (gauge generators, structure constants, zeromodes, $\ldots$) as coefficients. In this sense the master equation generates the gauge structure of the theory. Hence, as said above, it provides a very compact way to write down the gauge structure of a theory.
\end{itemize}
In \cite{Coomans:2010xd} a generic form for the generalized action $S$ is proposed
\begin{eqnarray}
S&=S_0&+\phi^*_iR^i{}_{a_0}c_{(0)}{}^{a_0}+c^*_{(0)}{}_{a_0}\left(Z_{(1)}{}^{a_0}{}_{a_1}c_{(1)}{}^{a_1}+\frac{1}{2}T^{a_0}{}_{b_0c_0}c_{(0)}^{c_0}c_{(0)}{}^{b_0}\right) \nonumber \\
     &&+\phi^*_i\phi^*_j\left(\frac{1}{2}V_{(1)}^{ji}{}_{a_1}c_{(1)}{}^{a_1}+\frac{1}{4}E^{ij}{}_{a_0b_0}c_{(0)}^{a_0}c_{(0)}{}^{b_0}\right)+\ldots,
\end{eqnarray}
with $S_0$ the action in ($\ref{action}$) and $R^i{}_{a_0}$, $T^{a_0}{}_{b_0c_0}$, $Z_{(1)}{}^{a_0}$, $\ldots$ some unspecified tensors. After imposing the master equation (\ref{masterequation}) it is shown that these tensors correspond respectively to the gauge generators, structure `constants', zeromodes, $\ldots$ that were discussed in the previous section.

\section{Conclusions}
Conventional gaugings break the underlying duality covariance of the theory. This is a problem when systematically classifying gauged supergravities. The embedding tensor formalism provides a method to circumvent this problem and to describe generalized gaugings while preserving duality covariance. However, it demands extra ingredients such as a tensor hierarchy and extra gauge transformations, which complicate the gauge structure of the theory. One argues that the gauge structure encountered in the embedding tensor formalism is soft, open and reducible. Hence, it is appropriate to reformulate it in terms of the field-antifield formalism. Once rewritten it becomes clear that the original classical hierarchy of the gauge fields also appears at the level of the ghosts, ghosts for ghosts, etc. Finally, it is found that the master equation is able to produce the complete gauge structure of the theory.

We thank the organizers of the XVIth European workshop in String Theory in Madrid for the offered opportunity to present this work, which is supported in part by the FWO - Vlaanderen, Project No. G.0235.05, and in part by the Federal Office for Scientific, Technical and Cultural Affairs through the IAP Programme P6/11-P.

\newpage
\bibliography{BVref}
\bibliographystyle{toine}
\end{document}